\documentstyle[12pt,bezier,epsfig]{article}
\def\theequation{\arabic{section}.\arabic{equation}}
\newcommand{\ezero}{\setcounter{equation}{0}}

\newcommand{\mr}{\mathrm}
\newcommand{\bq}{\begin{equation}}
\newcommand{\eq}{\end{equation}}
\newcommand{\ba}{\begin{eqnarray}}
\newcommand{\ea}{\end{eqnarray}}
\newcommand{\ds}{\displaystyle}

\newcommand {\nll}{\nonumber \\}
\newcommand {\nn} {\noindent}
\newcommand {\lm} {\mbox{${\mr L}_{\mathrm m }$}}

\newcommand {\Ql} {\mbox{$Q^2_{l}    $}}
\newcommand {\QlS}{\mbox{$Q^4_{l}    $}}
\newcommand {\Qh} {\mbox{$Q^2_{h}    $}}

\newcommand {\yl} {\mbox{$y  _{l}    $}}
\newcommand {\xl} {\mbox{$x  _{l}    $}}
\newcommand {\yll}{\mbox{$y  _{l_1}  $}}

\newcommand {\yh} {\mbox{$y  _{h}    $}}

\newcommand {\ylpl}{\mbox{$Y  _{+}   $}}

\newcommand {\AML}{\mbox{$m^2$}}

\newcommand {\COSPHI}{\mbox{$\cos{\varphi}_{\gamma}$}}

 \newcommand{\EPR  }{\mbox{$E_p$}}
 \newcommand{\PPR  }{\mbox{$p_p$}}
 \newcommand{\EEL  }{\mbox{$E_e$}}
 \newcommand{\PEL  }{\mbox{$p_e$}}

 \newcommand{\EELP }{\mbox{$E^{'}_e$}}
 \newcommand{\PELP }{\mbox{$p^{'}_e$}}
 \newcommand{\EGAMMA}{\mbox{$E_{\gamma}         $}}
 \newcommand{\PHI   }{\mbox{$\varphi_{\gamma}   $}}
 \newcommand{\COSTE }{\mbox{$\cos\theta^{'}_e $}}
 \newcommand{\SINTE }{\mbox{$\sin\theta^{'}_e $}}
 \newcommand{\SINTG }{\mbox{$\sin\theta_{\gamma}$}}
 \newcommand{\COSTG }{\mbox{$\cos\theta_{\gamma}$}}
 \newcommand{\Qlll  }{\mbox{$Q^4_l            $}}
 \newcommand{\Qhll  }{\mbox{$Q^4_h            $}}
 \newcommand{\TT    }{\mbox{$\ds S y_h        $}}
 \newcommand{\AMP   }{\mbox{$M^2              $}}

\newcommand{\LL    }{\mbox{${\left(\ln\frac{\Ql}{\ds m^2}-1\right)}$}}
\newcommand{\TPHCOB}{\mbox{${4 \pi \alpha^2}$}}

\newcommand{\EGABAR}{\mbox{${{\bar E}_{\gamma}}$}}

\newcommand{\DELTAVR}{\mbox{$\delta_{_{\rm{VR}}}$}}

\begin{document}

 \psfull

\thispagestyle{empty}
\onecolumn
\begin{flushleft}
DESY 96--213
\\
hep-ph/9612203
\\
November 1996
\end{flushleft}

\noindent
\vspace*{0.50cm}
\begin{center}
{ \huge 
Deep Inelastic Scattering 
\\
with Tagged Photons at HERA

\vspace*{3.4cm}
     } 
\nn
{\Large
D. Bardin$\;^{a,b}$,$\;$
L. Kalinovskaya$\;^{b,}$\footnote{Supported by the
  Heisenberg-Landau Program},
and
T. Riemann$\;^{a}$}
\\
\vspace*{1.5cm}
\end{center}

\large
$^a$ 
Deutsches Elektronensynchrotron DESY, 
Institut f\"ur Hoch\-ener\-gie\-phy\-sik,
  Platanenallee 6, D-15738 Zeuthen, Germany

\bigskip

$^b$
Bogoliubov Laboratory for Theoretical Physics, 
Joint Institute for Nuclear Research,
ul. Joliot-Curie 6, RU-141980 Dubna, Russia

\vfill

\centerline{\large Abstract}

\bigskip

\normalsize
\noindent
We calculate cross-sections for neutral current deep inelastic
scattering at HERA with photon tagging.
Both, the exact lowest-order cross-section and a leading logarithmic
approximation of next-order corrections are calculated.
The latter amounts to less than 20\% in a large kinematical range. 
The integrations are performed numerically but without relying on
Monte Carlo methods.

\vfill

\newpage

\section{Introduction
\label{intro}
}
\ezero
We calculate cross-section predictions for the reaction
\ba
e(k_1) + p(p_1) \longrightarrow e(k_2) + X(p_2) + \gamma(k),
\label{eq1}
\ea
where the photon is assumed to be observed (tagged).
The calculated cross-section will be differential in energy $E_e'$ and
angle $\theta_e'$ of the final state electron and in the energy
$E_{\gamma}^{\rm{vis}}$ of the photon.
Equivalently, the usual scaling variables $x, y$ together with
$E_{\gamma}^{\rm{vis}}$ may be chosen\footnote{
We note that the usual definition of $x$ and $y$ in terms of leptonic
variables is used. They are not recalculated using the reduced
electron beam energy after emission of the observed photon. 
}.
The angular range covered by the photon tagging device is taken into
account by kinematical cuts in the course of integrations. 
In the numerical results we will follow the experimental set-up of the H1
collaboration with the photon tagging device located in the forward
direction with respect to the electron beam\footnote{ 
This direction is called backward
direction [with respect to the proton beam] in the H1 collaboration.
}.
The lowest-order cross-section is determined without approximations.
Since one expects large photonic corrections in some kinematical
ranges, we tried to estimate them with a leading logarithmic
approximation. 
What we do is certainly not a rigorous approach and cannot replace a
complete second-order calculation.
In this connection the neglection of effects due to photon
interferences has to be mentioned.
They are of next-to-leading order, but may be non-negligible for a
forward tagging device.
For the time being, we hope to cover the most substantial corrections.  

In Section~\ref{low}, the lowest-order cross-section is calculated,
Section~\ref{I} contains a discussion of the inclusive cross-section.
The leading-logarithmic corrections are calculated in Section~\ref{lla} and 
numerical results are presented in Section~\ref{discussion}. 
Some technical details may be found in the Appendix. 
\section{The Lowest-Order Cross-Section
\label{low}
}
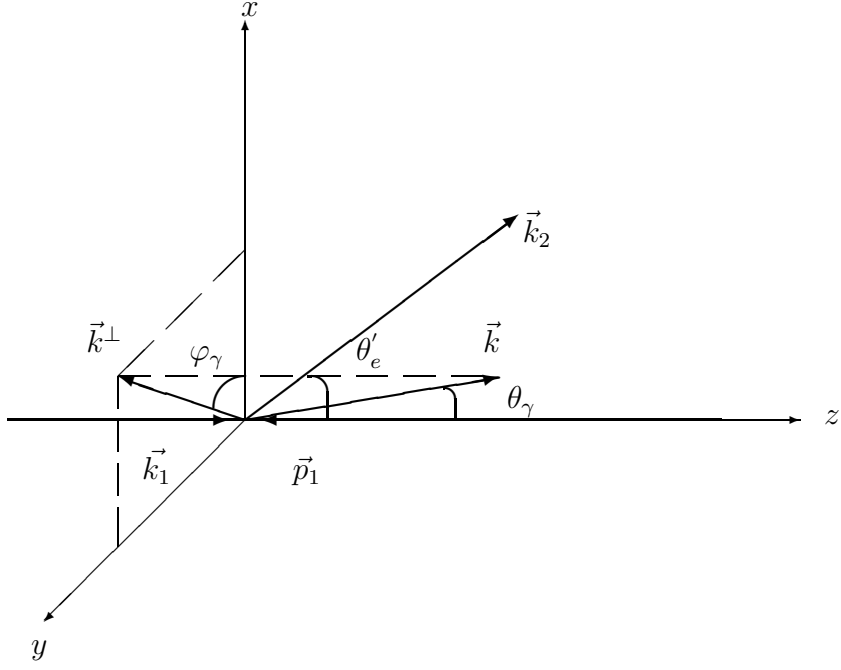
\begin{figure}[thbp]
\begin{minipage}[bhtp]{16.0cm}{
\begin{center}
 \begin{picture}(300,150)(-90,+60)
\setlength{\unitlength}{3pt}
\thicklines
\put(-7,30){${\varphi_{\gamma}}$}             %
\put(00,22.5){\line(-3, 1){15}}             %
\put(00,22.5){\line( 6, 1){32}}             %
\put(32,28  ){\vector(4 , 1){0}}            %
\put(-16,28 ){\vector(-3, 1){0}}            %
\put(-13,15 ){$ \vec {k_1}   $}             
\put(00,22.5){\line(1,0){60}}               
\put(2,22.5){\vector(-1, 0){0}}             %
\put(6,15){${\vec p_1}$}                    %
\put(00,22.5){\line(-1,0){30}}              
\put(-2,22.5){\vector(1, 0){0}}             %
 \put(0,22.5){\line(4, 3){34}}              
 \put(35,45){${\vec k}_2$}                  
 \put(34.5,48.5  ){\vector(4, 3){0}}        %
\thinlines
  \put(-16,28){\line  (1,1){4.5}}
  \put(-10,34){\line  (1,1){4.5}}
  \put(-4 ,40){\line  (1,1){4}}
\multiput(-16,28)(6,0){8 }{\line(1,0){4}} 
\put(-16,28  ){\line(0,-1){4} }
\put(-16,22.5){\line(0,-1){4} }
\put(-16,17  ){\line(0,-1){4} }
\put(-16,11.5){\line(0,-1){5} }
\put(00,22.5){\line(0,1){50} }             
\put(-.5,73.5){$ x $}                      
\put(00,73 ){\vector(0, 1){0}}             %
\put( 00,22.5){\line  (-1,-1){25.3}}  
\put(-25.4,-3){\vector(-1,-1){0 }}    %
\put(-27  ,-7){$ y $}                 %
\put(73,22   ){$ z $}                 
\put(60,22.5){\line  (1,0){10}}       %
\put(70,22.5){\vector(1,0){0} }       %
\thicklines
\put(0 ,24  ){\oval( 8,8 )[tl]}             %
\put(25   ,22.5){\oval(3  ,8  )[tr]}        
\put(7.9  ,22.5){\oval(5  ,11 )[tr]}        
\put(14,30  ){${\theta^{'}_e}     $}        %
\put(33,24  ){${\theta    _{\gamma}}$}        %
\put(30,31  ){${\vec k      }$}             
\put(-20,31 ){${\vec k}^{\perp}$}           
 \end{picture}
\end{center}
}\end{minipage}
\vspace*{3.cm}
\caption{\it Three-momenta in the HERA frame
\label{fig1}
}
\end{figure}

\ezero
In this section we derive a set of formulae, which allow the lowest-order
description of deep-inelastic cross-sections with tagged photons, where the
latter may have arbitrary kinematics.
Obviously, the treatment of arbitrary cuts in energy and production
angles is possible only with numerical methods.
Thus, we decided to tackle the problem not with the semi-analytical
approach as 
developed in our previous papers.
Instead, we will numerically integrate the squared matrix element over
the phase space 
imposing the appropriate cuts explicitly.
For this aim, we use natural variables in the HERA system.
Besides the four-momenta of initial state electron and proton, this
are variables, which are
directly measured by the apparatus: angles and energies of final state 
electron, $\EELP, \theta_e'$, and of the photon, $\EGAMMA,
\theta_{\gamma}, \varphi_{\gamma}$.
The following representations for the 4-vectors $p_1, k_1, k_2, k$ have been
chosen (see also Figure~\ref{fig1}):
\ba
 p_1 &=& (E_p,0,0,-p_p),         
\\ \nll 
 k_1 &=& (E_e,0,0, p_e),         
\ea

\bigskip

\ba
 k_2 &=& (E^{'}_e, p^{'}_e \sin\theta^{'}_e,0,p^{'}_e \cos \theta^{'}_e), 
\\ \nll
 k   &=& E_{\gamma} (1,\sin\theta_{\gamma} \cos\varphi_{\gamma},
                      \sin\theta_{\gamma} \sin\varphi_{\gamma},
                      \cos\theta_{\gamma}), 
\\ \nll
 p_2 &=& p_1+k_1-k_2-k.
\ea

In these variables, the invariants $s, Q_l^2, y_l$ are:
  \ba
   s &=& (k_1+p_1)^2 
= (E_e+E_p)^2-(p_e-p_p)^2
= S+M^2+m^2,  
\label{s}
 \\ \nll
S &=&  2p_1k_1
= 2(E_eE_p +p_e p_p),
\label{S}
\\ \nll  
  y_l &=&
 \frac{p_1(k_1-k_2)}{p_1k_1}
= 1-2(\EPR\EELP+\PPR\PELP\COSTE)/S,
\label{yl}
\\ \nll
 Q^2_l
&=& - (k_1-k_2)^2
=2(\EEL\EELP-\PEL\PELP\COSTE)-2m^2.
\label{QlQl}
\ea
With $m$ and $M$ we denote the electron and proton masses, respectively.
We will also use the $\pi$ production threshold,
\ba
   {\bar M} &=&  M + m_{\pi}, 
\label{Mbar}
\ea 
which is the lowest mass of the inelastic final hadronic system.

The angular variables are unlimited, 
\ba
-1 \leq&  \COSTE& \leq +1,
\label{lim_coste}
\\ \nll
-1 \leq&  \COSTG& \leq  +1,
\label{lim_costg}
\\ \nll
0  \leq & \varphi_{\gamma}& \leq  2\pi ,
\label{lim_phi}
\ea
while the allowed ranges of electron and
photon energies have to be determined:
\ba
m  \leq& \EELP& \leq  E^{{'}\max}_e(\theta^{'}_e) ,
\label{Eepr}
\\ \nll
0  \leq& E_{\gamma}& \leq 
E^{\max}_{\gamma}(E_e^{\prime},\theta^{'}_e,\theta_{\gamma},\varphi_{\gamma})
.  
\label{Egam} 
\ea
The upper limits are derived in Appendix~\ref{appA}: 
\ba
E^{{'}\max}_e   = \frac{1}{ S_\theta}
    \left[( E_e + E_p)(E_e E_p + p_e p_p + m^2 - \Delta M^2/2 )
                  + (p_e - p_p) \cos\theta^{'}_e \sqrt{L_\theta}
                \right] ,
\label{starl} 
\ea
with
\ba
 L_\theta   &=&   (E_e E_p +p_e p_p +m^2- \frac{1}{2}\Delta M^2)^2-m^2 
S_\theta, 
\\ \nll
 S_\theta   &=&   ( E_e + E_p)^2-( p_e-p_p)^2 \cos^2\theta^{'}_e ,
\\ \nll
 \Delta M^2 &=& {\bar M}^2 - M^2,
\ea
and 
\ba
E^{\max}_{\gamma} = \frac
           {S+2m^2+M^2-{\bar M}^2 -2(E_e+E_p) E^{'}_e
+2(p_e-p_p)p^{'}_e \cos\theta^{'}_e}
           {2\left[E_e +E_p 
-E^{'}_e+(p^{'}_e\cos\theta^{'}_e-[p_e-p_p])\cos\theta_{\gamma} 
                   +p^{'}_e\sin\theta^{'}_e
                           \sin\theta_{\gamma}
                           \cos\varphi_{\gamma}\right]} .
\label{emaxg}
\ea

The phase space is defined as follows:
\ba
d\Gamma &=& \frac{d\vec{k}_2}{2k_2^0} \frac{d\vec{k}}{2k^0}
    \frac{d\vec{p}_2}{2p_2^0} dM_h^2 \,
    \delta^{4}(k_1+p_1-k_2-k-p_2),
\label{gamma1}
\ea
with
\ba
M_h^2 &=&  (p_1 + k_1 -k_2 - k)^2.
\label{Mh}
\ea
We show in appendix~\ref{appA} that the 
doubly-differential phase space 
in terms of natural variables takes the following form:
\ba
\frac{ d\Gamma                  }
     { p^{'}_e d\EELP d \COSTE  }
 = \frac{\pi }{2}
 \int_0^{2\pi}d\PHI \int_{-1}^{+1} d\COSTG  \int_{0}^{E^{\max}_{\gamma}}
 E_{\gamma} dE_{\gamma}.
\label{dgammall}
\ea

The totally differential cross-section in natural variables is:
\ba
 \frac{d^5 \sigma_{\rm{brem}}}
{d\COSTE d\EELP d\COSTG d\varphi_{\gamma} d E_{\gamma}} 
  &=& \frac{4\alpha^3}{\pi S} \PELP E_{\gamma} \frac{1}{Q^4_h}
                            \sum^3_{i=1} S_{i} {\cal F}_i.
\label{032}
\ea
The functions ${\cal F}_i$ are generalized structure functions.
Explicit expressions for them may be found in
Equations~(I.2.18)~\footnote{ 
Here and henceforth, we mark equations taken from
reference~\cite{MI} with I.}.

The totally differential `radiator functions' $S_{i}$ for the
radiative process 
are, with account of minor improvements and modifications, taken from 
Equations~(I.2.40)--(I.2.42): 
\ba
 S_{1} &=& - 2\AML \Qh \left(\frac{1}{z_1^2}+\frac{1}{z_2^2}\right)
         + \frac{\Qlll+\Qhll}{z_1 z_2} + 2,
\\ \nll
S_{2}  &=& \frac{1}{\TT}   \Biggl\{
              - 2  \AML \left[\frac{S^2\yll (\yll+\yh)-\AMP\Qh}{z_1^2}
              + \frac{S^2(1  -\yh)-\AMP \Qh}{z_2^2} \right]        
 \nll
         & &  + \frac{S^2 \Qh (\ylpl-\yl\yh)-\AMP(\Qlll+\Qhll)}{z_1 z_2}
              - S^2 \yh \left(\frac{1}{z_1}+\frac{\yll}{z_2}\right)
              - 2 \AMP       \Biggr\},
\\ \nll  
S_{3}  &=&   \frac{1}{\yh} \Biggl\{
             - \AML \Qh \left[ \frac{2\yll+\yh}{z_1^2}+\frac{2-\yh}{z_2^2}
\right] 
             + \frac{ \Ql\Qh (2-\yl)}{z_1 z_2}
 \nll
         & & - \Qh \left(\frac{\yll}{z_1}-\frac{1}{z_2} \right)
             - \yh \frac{\Ql+\Qh}{2} \left(\frac{1}{z_1}+\frac{1}{z_2} \right)
                    \Biggr\}.
\ea
Above, we retained $m$ where needed and $M$ throughout.
The following abbreviations and definitions have been used:
\ba
     \yll  &=& 1-\yl,          
\\ \nll
     \ylpl &=& 1+y^2_{l_1},    
\\ \nll
  \yh      &=& \frac{p_1(p_2-p_1)}{p_1k_1}
            =
   \yl-\frac{2 \EGAMMA}{S}(\EPR +\PPR \COSTG),     
\\ \nll
  z_1      &=& 2k_1k
            = 2 \EGAMMA \left( \EEL - \PEL \COSTG  \right),     
\\ \nll
  z_2      &=& 2k_2k  = z_1 + \Ql - \Qh 
\nll
           &=& 2 \EGAMMA \left[ \EELP- \PELP
\left(\COSTE\COSTG+\SINTE\SINTG\COSPHI\right)\right],
\ea
\ba
 \Qh &=& - (p_2-p_1)^2
\nll
&=&
\Ql      -2\EGAMMA \left[ (\PEL-\PELP \COSTE) \COSTG
                  -  \PELP    \SINTE  \SINTG \COSPHI
                  +  \EELP  -\EEL  \right].
\label{qh2}
\ea
The definitions reflect the radiative kinematics:
the difference $\Qh-\Ql$ as well as $z_1$ and $z_2$
are proportional to $\EGAMMA$. In Appendix~\ref{appA}, the same is shown for
$W^2-M_h^2$. 

\bigskip

{From} the above expressions, any radiative cross-section, if measured
in natural variables, may be calculated by simply imposing the
corresponding cuts when integrating~(\ref{032}). 

The lowest-order radiative cross-section becomes: 
\ba
 \frac{d^3 \sigma_{\rm{brem}}}
{d\COSTE d\EELP d E_{\gamma}} 
=
\frac{4\alpha^3}{\pi S} \PELP
\int  d\COSTG 
 \int d\PHI 
\frac{E_{\gamma}}{Q^4_h}
                            \sum^3_{i=1} S_{i} {\cal F}_i,
\label{lowest}
\ea
with reasonably defined boundaries for the angular integrations.
In case of a forward tagging device as being used by the H1 and ZEUS 
collaborations:
\ba
\cos\theta_{\gamma} \in \left( 1- \frac{1}{2}\Delta\theta_{\gamma}^2,
1 \right)
, \qquad \varphi_{\gamma} \in (0, 2\pi) .
\label{lowestl}
\ea
Here, $\Delta\theta_{\gamma}$ describes the opening angle of the
tagging device\footnote{
Strictly speaking, one should take into account the actual form of the
tagger (being e.g. rectangular rather than circular); we leave this as an
exercise to the reader.}. 

For applications, it may be useful finally to change variables: 
 \ba
 \frac{d^3 \sigma_{\rm{brem}}}
{d\xl     d\yl dE_{\gamma}  } 
= 
{\cal J }
 \frac{d^3 \sigma_{\rm{brem}}}
{d\COSTE d\EELP dE_{\gamma} },
\label{bb}
 \ea
with $Q_l^2 = x_l y_l S$.
The Jacobean is easily determined from~(\ref{yl}) and~(\ref{QlQl}):
\ba
{\cal J} =
  \left| \frac  {\partial( \COSTE, \EELP)}
                {\partial( \xl  ,   \yl )}
  \right|
  =\frac  {\yl S  }
          {2     \PELP }.
 \ea

\begin{figure}
\begin{center}
\mbox{\epsfig{file=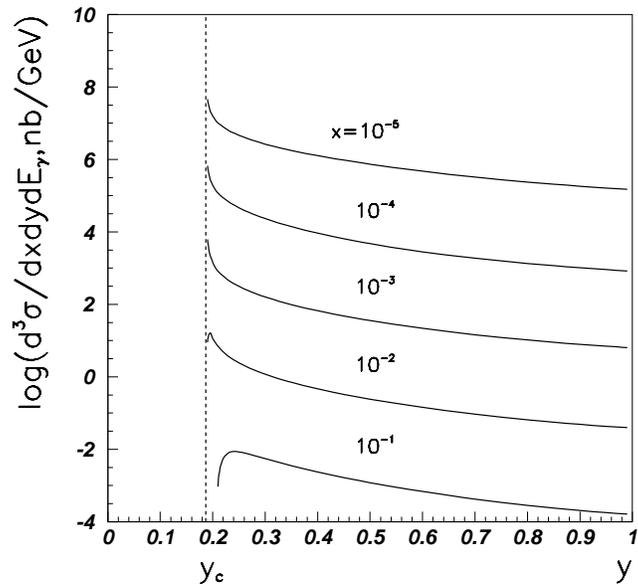,height=9cm,width=9cm}}
\end{center}
\caption{
\it
Lowest-order cross-section for HERA kinematics with $E_{\gamma}=5$ GeV 
and $\Delta\theta_{\gamma} = 0.5$ mrad
\label{Flow1}
}
\end{figure}

\begin{figure}
\begin{center}
\mbox{\epsfig{file=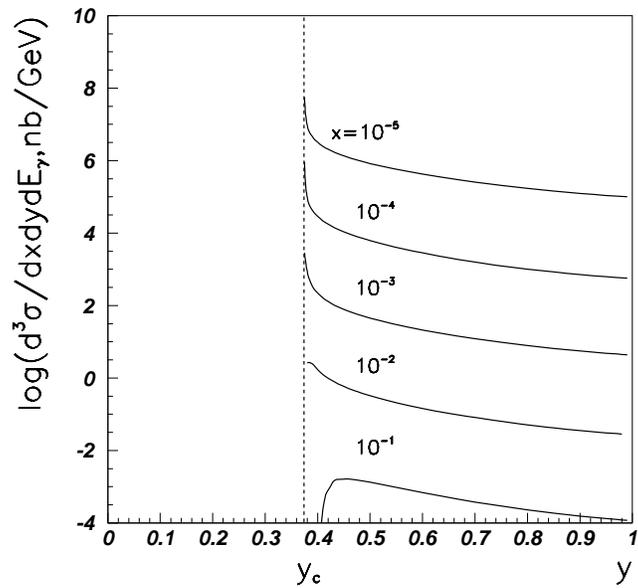,height=9cm,width=9cm}}
\end{center}
\caption{
\it
The same as Figure~2 
but with $E_{\gamma}=10$ GeV
\label{Flow}
}
\end{figure}

We show the lowest-order cross-sections (\ref{bb}) in 
Figures~\ref{Flow1} and~\ref{Flow}
for $E_{\gamma}=5,10$ GeV and $\Delta\theta_{\gamma} = 0.5$ mrad.
The updated Fortran program {\tt HECTOR}~\cite{hector} with flag
settings {\tt ISSE}=1 and 
{\tt ISCH}=0 (selecting structure functions {\tt CTEQ3(LO)}) is used for all
numerical results.  
As is seen in the figures, the minimal value of $y$ depends on
both $x$ and $E_{\gamma}^{\rm{vis}} = E_{\gamma}$.  
In addition, it depends on the angular cuts.
For a forward tagging device, its absolute minimum may be approximated
by 
\ba
y_c = \frac{ E_{\gamma}^{\rm{vis}}}{E_e}.
\label{yc}
\ea
The cross-section resembles the typical behavior of inclusive deep
inelastic scattering.

\section{The Inclusive Cross-Section 
\label{I}
}
\ezero
In the preceding Section, we made the assumption that the energy of the tagged
photon will be observed.
Events with vanishing photon energy were automatically excluded 
and thus all the problems related to the infrared singularity.
In this Section, we will treat the photon energy as inclusive variable,
i.e. determine the doubly-differential cross-section
$d^2\sigma/d\COSTE d\EELP$ as the sum of Born cross-section, vertex correction,
and real brems\-strah\-lung contributions:
\ba
 \frac{d^2\sigma}
      {d\cos\theta^{'}_e dE^{'}_e}  
&=&
\frac{d^2\sigma_{\rm{Born}}}{d\cos\theta^{'}_e dE^{'}_e}
\left( 1 + \delta_{\mr{vert}} \right)
+  \int d E_{\gamma}
      \frac{d^3\sigma_{\rm{brem}}}
{dE_{\gamma} d\cos\theta^{'}_e dE^{'}_e} .
\label{eq_bb0}
\ea
The purpose is two-fold. 
We want to have the opportunity to perform a
numerical comparison of the present, deterministic calculations with
those performed in the semi-analytical approach; and the latter are
doubly-differential.
Further, in the next Section we estimate higher-order QED
corrections using the leading-logarithmic approximation for them. 
In doing so, we will have to make use of~(\ref{lowest}) without a
restriction on the lower limit of $E_{\gamma}$. 

We will consider~(\ref{lowest}) without cuts on the photonic angles, 
\ba
 \frac{d^3 \sigma_{\rm{brem}}}
{d\COSTE d\EELP d E_{\gamma}} 
=
\frac{4\alpha^3}{\pi S} \PELP
\int _ {-1} ^ {1} d\COSTG 
 \int_0^{2\pi}d\PHI 
\frac{E_{\gamma}}{Q^4_h}
                            \sum^3_{i=1} S_{i} {\cal F}_i,
\label{n_g}
\ea
and derive the lowest-order QED corrections
$d^2 \sigma_{_{\rm{QED}}} / d\xl d\yl $ to
the cross-section 
 \ba
 \frac{d^2 \sigma_{\rm{theor}}}{d\xl d\yl } = \frac{d^2
   \sigma_{\rm{Born}}}{d\xl d\yl } 
+
\frac{d^2 \sigma_{_{\rm{QED}}}}{d\xl d\yl }.
 \label{micrs}
 \ea
Then, $d^2\sigma_{_{\rm{QED}}}$ has to be compared to the semi-analytical
result, see e.g. Equations~(I.4.40) and~(I.5.3). 
For this aim, we have to multiply~(\ref{n_g}) by the 
Jacobean~(\ref{jacob}), 
then to integrate over $E_{\gamma}$, and therefore to regularize the
infrared divergence.
{From}~(\ref{eq_bb0}), the  $ {d^2 \sigma_{_{\rm{QED}}}} $ is to be
calculated now: 
\ba
 \frac{d^2\sigma_{_{\rm QED}}}
      {d\cos\theta^{'}_e dE^{'}_e}  &=&
       \int^{+1  }_{-1} d\cos\theta_{\gamma}
       \int^{2\pi}_{0}  d\varphi_{\gamma}
       \int^{E^{\max}_{\gamma}}_0  d E_{\gamma}
\Biggl\{
      \frac{d^5\sigma_{\rm{brem}}}
{d\cos\theta^{'}_e dE^{'}_e d\cos\theta_{\gamma} d\varphi_{\gamma}
  dE_{\gamma}} 
       \theta(E_{\gamma} - {\bar E}_{\gamma})
\nll
  &&  +    \frac{1}{\Delta \Omega_{\gamma}}
      \frac{d^2\sigma_{\rm{Born}}}{d\cos\theta^{'}_e dE^{'}_e}
\left[ 1 + \delta_{_{\mr{VR}}} ({\bar E}_{\gamma}) \right] 
\delta(E_{\gamma}) 
\Biggr\},
\label{eq_bb}
\ea
with a cutoff ${\bar E}_{\gamma}$ on the minimum photon energy for the
numerical integration. 
The correction $\delta_{_{\mr{VR}}}$ will be discussed below.
Further, it is 
\ba
\Delta \Omega_{\gamma} = 
       \int d\cos\theta_{\gamma}
       \int d\varphi_{\gamma}.
\label{om}
\ea
In~(\ref{eq_bb}), the angular integrations are unrestricted, $\Delta
\Omega_{\gamma} = 4 \pi$ and $d^5\sigma_{\rm{brem}}$
 is a completely differential, infrared-finite bremsstrahlung cross-section
and may be treated numerically as it stands.
It is the analogue to Equation~(I.5.3) without the subtracted terms.

In fact, we have to follow a different strategy for the solution of
the soft-photon problem. In the semi-analytical approach, we performed
a subtraction of the infrared divergent part of the integrand thus
making the numerical integration infrared finite in the limit
$E_{\gamma} \to 0$.
This subtraction was compensated by an analytical integration of the
subtracted part.
We specially had to choose a reference frame where the limits of
photon energy variation 
are independent of the angular variables, this way considerably
simplifying the integrations.
This was possible since the net cross-section was defined in
a covariant way.
Here, the situation is quite different. 
We have to use the above defined set of natural variables in order to
finally be allowed to apply the cuts.
As was discussed, the photon energy varies in limits which depend in a
complicated way on the photon angles.
It is important for the calculation that one may
choose $\EGABAR$ so small that
it is well inside a region where $E_{\gamma}$ {\em is unrestricted} by
the photon angles.
That this region exists for arbitrary values of $\theta^{'}_e,
E^{'}_e$ may be seen in Figure~\ref{3} where the allowed region
for photon energies is shown below the shaded surface as function of
these angles.
At $\cos \theta_{\gamma}=0$, the minimal $E_{\gamma}^{\max}$ is
reached.
If ${\bar E}_{\gamma}$ is smaller, there is no influence of the photon
angles on the photon energy boundary. We have to choose:
\ba
{\bar E}_{\gamma} < 
\frac
           {S+2m^2+M^2-{\bar M}^2 -2(E_e+E_p) E^{'}_e
+2(p_e-p_p)p^{'}_e \cos\theta^{'}_e}
           {2\left[E_e +E_p 
-E^{'}_e+(p^{'}_e\cos\theta^{'}_e-[p_e-p_p])
\right]} .
\label{emaxb}
\ea
We split the correction into pieces as it was done in~(I.4.40):
\ba
\frac{d^{2}{\sigma}_{_{\mr{QED}}}}
{d\cos\theta_e^{\prime} d E_e'}
&=&
\frac{\alpha}{\pi}\;\delta_{\mr {vert}}
\frac {d^2 \sigma_{\mr Born }} {d\cos\theta_e^{\prime} d E_e'} 
+
\frac {d^2 \sigma_{\mr{soft}}^{\mr{IR}}} {d\cos\theta_e^{\prime} d E_e'} 
+
\frac {d^2 \sigma_{\mr{hard}}^{\mr{IR}}} {d\cos\theta_e^{\prime} d E_e'} 
+
\frac {d^2 \sigma_{\mr{R}}^{\mr{F}}} {d\cos\theta_e^{\prime} d E_e'}
\nll
&\equiv&
\frac{\alpha}{\pi}\;\delta_{_{\mr {VR}}}({\bar E}_{\gamma})
\frac{d^{2} {\sigma}_{\mr Born}}    
{d\cos\theta_e^{\prime} d E_e'}
+\frac{d^{2}{\sigma}_{\mr R}^{\mr F}}
    {d\cos\theta_e^{\prime} d E_e'},
\label{62b}
\ea
with
\ba
    \delta_{_{\mr {VR}}} ({\bar E}_{\gamma})
&=&
 \delta_{\mr {vert}}(\mu)
+\delta^{\mr {IR}}_{\mr {soft}}(\epsilon,\mu)
+\delta^{\mr {IR}}_{\mr {hard}}(\epsilon,{\bar E}_{\gamma})
.
\label{64c}
\ea
Here, ${d^2 \sigma_{\mr{R}}^{\mr{F}}} /  {d\cos\theta_e^{\prime} d
  E_e'}$ is the result of the numerical integration of the first term
in~(\ref{eq_bb}) and the Born cross-section is: 
\ba
      \frac{d^2\sigma_{\rm{Born}}}{d\cos\theta^{'}_e dE^{'}_e}
                  &=& \frac{\TPHCOB}{S} \frac{\PELP}{\QlS}
      \sum_{i=1}^3 S^{^{\rm B}}_{i} {\cal F}_i,
\ea
with the Born radiators
\ba
S^{^{\rm B}}_{1} &=&2\Ql, 
\\ \nll
S^{^{\rm B}}_{2} &=&\frac{2}{S\yl}\left[(1-\yl)S^2-M^2\Ql\right], 
\\ \nll
S^{^{\rm B}}_{3} &=&\frac{2-\yl}{\yl}\Ql .
\ea
The hard part of the infrared divergent corrections
is defined by~Equation~(I.4.8):
 \ba
      \delta_{\rm{hard}}^{\mr{IR}}(\epsilon,\EGABAR)
&=&
   \frac{1}{\pi}
\int_{\epsilon}^{\EGABAR} d E_{\gamma} E_{\gamma} \int_{-1}^{1} d \cos
\theta_{\gamma} \int_{0}^{2\pi} d \varphi_{\gamma} \, {\cal F}^{\rm{IR}}       
=
 2\ln \frac{\EGABAR}{\epsilon} \, \left(\ln \frac{Q_l^2}{m^2}-1 \right),
\nll
 \ea
where ${\cal F}^{\rm{IR}}$ is the Low factor (see e.g. Equation~(I.4.5)):
\ba
{\cal F}^{\rm{IR}}
=
\frac{Q^2}{z_1z_2} - m^2\left(
\frac{1}{z_1^2} + \frac{1}{z_2^2}\right) .
\label{ir3}
\ea
The integrals used are given in the Appendix.

\bigskip

The infrared divergent part of the soft photon contribution is: 
 \ba
      \delta_{\mr{soft}}^{\rm{IR}}(\epsilon,\mu)
&=&
   \frac{1}{\pi}
\int_{0}^{\epsilon} d E_{\gamma} E_{\gamma} \int_{-1}^{1} d \cos
\theta_{\gamma} \int_{0}^{2\pi} d \varphi_{\gamma} \, {\cal F}^{\rm{IR}} .      
\label{sir}
\ea
With its calculation, one may follow Reference~I until reaching
Equation~(I.4.26): 
\ba
\delta^{\mr {\rm{IR}}}_{\mr {soft}}(\epsilon,\mu)
= 2\Bigl[ {\cal P}^{\mr {IR}}+\ln \frac{2\epsilon}{\mu} \Bigr]
\left[ (Q_l^2+2m^{2}) \lm        -1 \right]
+~
 {\cal S}_{\Phi}
 + \frac{1}{2 \beta_{1}} 
  \mbox{ln}\frac{1+\beta_1}{1-\beta_1}
 +\frac{1}{2 \beta_{2}} 
  \mbox{ln}\frac{1+\beta_2}{1-\beta_2}
,
\nll
\label{eqn415}
\ea
with
\ba
\lm &=& \frac{1}{\sqrt{\lambda_m}} \ln \frac{\sqrt{\lambda_m} + Q_l^2}
{\sqrt{\lambda_m} - Q_l^2},
\label{lmlamm}
\\ \nll
{\lambda}_{m}
&=&
Q_l^2 (Q_l^2+4m^2).
\ea
Calculating~(\ref{eqn415}) in the ultrarelativistic limit, 
with $\beta_1$ and $\beta_2$ being the velocities of the initial and final
state electrons in the HERA frame,
the infrared divergent part gets:
 \ba
        \delta_{\mr{soft}}^{\rm{IR}}(\epsilon,\mu) &=&
       2 \left( P^{\rm{IR}} + \ln \frac{2\epsilon}{\mu}\right)
        \left( \ln \frac{Q_l^2}{m^2}-1 \right)
       +\ln\frac{4\EEL\EELP }{m^2}
       +{\cal S}_{\Phi} .
\ea
The function ${\cal S}_{\Phi}$ also reflects the dependence of the soft
photon treatment on the reference system in which the corrections 
$\delta_{\mr{soft}}^{\rm{IR}}$ and $\delta_{\rm{hard}}^{\rm{IR}}$
are
calculated.
We calculate ${\cal S}_{\Phi}$ numerically using the integral
representation: 
\ba
{\cal S}_{\Phi}
&=& \frac{1}{2} \left(Q^2+2m^2\right)
 \int_{0}^{1} \frac {d\alpha}{ \beta_{\alpha}k_{\alpha}^2} \;
       {\ln}{ \frac{1-\beta_{\alpha}}{1+\beta_{\alpha}} },
\label{soft3}
\ea
with
\ba
 k_\alpha &=& k_1 \alpha +  k_2 (1-\alpha),
\\ \nll
   \beta_\alpha &=& \frac{|\vec k_\alpha|}{ k^0_\alpha},
\\ \nll
 k^2_\alpha &=& \Ql\alpha(1- \alpha) +  m^2,
\\ \nll
 |\vec k_\alpha|^2
    &=& p^2_e \alpha^2
      +      ( p^{'}_e)^2  (1-\alpha)^2
      + 2 p_e p^{'}_e \cos\theta^{'}_e \alpha (1-\alpha),
\\ \nll
  k^0_\alpha &=& \EEL \alpha+\EELP (1-\alpha).
\ea

The infrared singularity is compensated by the
contribution from the QED vertex correction (see e.g. Equation~(I.4.34)): 
 \ba
      \delta_{\rm{vert}}(\mu)=
       -2  \left( P^{\rm{IR}} + \ln \frac{m}{\mu}\right)
         \left( \ln \frac{Q_l^2}{m^2}-1 \right)
       - \frac{1}{2} \ln^2\frac{\Ql}{m^2}
       + \frac{3}{2} \ln \frac{Q_l^2}{m^2} - 2 + \frac{\pi^2}{6}. 
\ea
In sum:
\ba
\DELTAVR(\EGABAR)
  &=& 2\ln\frac{2\EGABAR}{m}\left( \ln \frac{Q_l^2}{m^2} -1\right)
       - \frac{1}{2} \ln^2\frac{\Ql}{m^2}
       + \frac{3}{2} \ln \frac{Q_l^2}{m^2}
      +\ln\frac{4E_e  E^{'}_e}{m^2}
    - 2 + \frac{\pi^2}{6}
      + {\cal S}_{\Phi}.
\nll
 \ea

In practice, we took ${\bar E}_{\gamma} = 10^{-4}$ GeV and verified
stability against variations.
With the {\tt Fortran} program {\tt HECTOR}, we further confirmed 
numerically with high precision that
cross-section~(\ref{62b}) agrees with that
calculated in the 
semi-analytical approach provided no cut is applied. 

\section{Higher-Order Initial-State Radiation 
\protect\newline
in Leading-Logarithmic
  Approximation
\label{lla}
} 
\ezero
So far we considered the lowest-order cross-section.
A calculation of higher-order photonic contributions is a non-trivial task
since the cross-section is not completely inclusive.
We will apply the leading-logarithmic approximation (LLA) which worked
amazingly well for inclusive cross-sections (see, e.g.,~\cite{MI}).

We begin with the generic LLA formula for photonic initial state corrections,
\ba
 \frac{d^2\sigma^{^{\rm{ini}}}} 
      {d\cos\theta^{'}_e dE^{'}_e}
  =  
     \int_0^1 dz \, \rho(z) \Biggl\{
      \frac{d^2\sigma_{_{\rm{QED}}}}
      {d\cos\theta^{'}_e dE^{'}_e}
    \Biggl|_{E_e \rightarrow z E_e}   \Biggr.
       \Theta_z
   -   \frac{d^2\sigma_{_{\rm{QED}}}}
       {d\cos\theta^{'}_e dE^{'}_e}
\Biggr\} .
\label{generic}
\ea
The totally inclusive ${\cal O}(\alpha)$ density is:
\ba 
\rho(z) = \frac{\alpha}{2\pi } \frac{1+z^2}{1-z} \LL .
\label{lho}
\ea
Only the initial electron energy scales, 
\ba
E_e \rightarrow {\hat E}_e = z E_e.
\label{esc}
\ea
The kinematical limits~(\ref{starl}) and~(\ref{emaxg}) of the final
state energies depend on the 
actual values of the angles and on ${\hat E}_e, E_p$.
Thus, we have to introduce a step function,
\ba
\Theta_z = \Theta({\hat W}^2 - M^2) \Theta({\hat M}_h^2 - M^2),
\ea
in order to ensure the non-trivial kinematical conditions~(\ref{w2})
and~(\ref{limit_eg}).
Further, we note that the Jacobean is unity in~(\ref{generic}) since this
differential cross-section is being calculated in terms of variables which
do not scale with $E_e \rightarrow z E_e$.

How is the cross-section~(\ref{generic}) related to an experiment with photon
tagging?
Formally, one has to `differentiate'~(\ref{generic}) with respect to
the visible photon energy.
But we have to care about the definition of $E^{\rm{vis}}_{\gamma}$.
And this depends largely on the specific experimental set-up.
There are two different cases to be considered:
\begin{itemize}
\item Case (i)
\\
The tagging device is {\em not in forward direction} of the
  electron beam;
\item Case (ii)
\\
It {\em is in forward direction}  of the
  electron beam.
\end{itemize}
The relevant object to be calculated is the three-fold differential
cross-section 
\ba
 \frac{d^3\sigma} { d\cos\theta^{'}_e dE^{'}_e dE^{\rm{vis}}_{\gamma}},
\ea
where the visible photon energy is the energy deposited in the 
tagging device.

We see two 
potential sources of inaccuracy which could make our calculations less reliable
than usual applications of LLA often are:
\begin{itemize}
\item
    For tagging in non-forward direction, the
assumption of inclusiveness of the process in the LLA photon angle is fulfilled
approximately while tagging in forward direction has to be treated
with care (see below).
\item
     The photon, which is treated with the LLA method and that of the Born
cross-section are identical particles but they are treated differently
and their interferences are neglected.
The treatment assumes them to be distinguishable.
The resulting inaccuracy, though, is of next-to-leading order.
\end{itemize}
\subsection{Case (i): Non-Forward Tagging Device
\label{nonf}
} 
In Case~(i), the visible photon energy is simply
\ba
E^{\rm{vis,(i)}}_{\gamma} = E_{\gamma}.
\ea
The LLA describes the emission of collinear photons, here from the
initial electron.
These forward photons cannot hit the non-forward tagging device and are thus
completely inclusive. 
Due to this, a direct application of~(\ref{generic})  is justified 
as a reasonable approximation and the neglect of identity of the photons is
expected to be not very influential.
The cross-section with account of leptonic initial-state LLA
corrections will be:
\ba
 \frac{d^3\sigma_{(i)}}
      {d\cos\theta^{'}_e dE^{'}_e d E^{\rm{vis}}_{\gamma}}
  =  
\frac{d^3\sigma_{\rm{brem}} }
      {d\cos\theta^{'}_e dE^{'}_e dE^{\rm{vis}}_{\gamma} }
+
 \frac{d^3\sigma_{(i)}^{^{\rm{ini}}}} 
      {d\cos\theta^{'}_e dE^{'}_e d E^{\rm{vis}}_{\gamma}},
\label{sigi}
\ea
with
\ba
 \frac{d^3\sigma_{(i)}^{^{\rm{ini}}}} 
      {d\cos\theta^{'}_e dE^{'}_e d E^{\rm{vis}}_{\gamma}}
&=&
\int_{\Delta\Omega_{\gamma}}
d\cos\theta_{\gamma}
d\varphi_{\gamma}
     \int_0^1 dz 
\rho(z)
\Biggl\{
      \frac{d^5\sigma_{\rm{brem}} }
      {d\cos\theta^{'}_e dE^{'}_e d\cos\theta_{\gamma}
        d\varphi_{\gamma} 
dE_{\gamma}
}
    \Biggl|_{E_e \rightarrow z E_e}   \Biggr.
       \Theta_z
\nll   
&& -~   \frac{d^5\sigma_{\rm{brem}}}
{d\cos\theta^{'}_e dE^{'}_e d\cos\theta_{\gamma} d\varphi_{\gamma}
  dE_{\gamma}} 
\Biggr\} .
\label{ini-i}
\ea

\subsection{Case (ii): Forward Tagging Device
\label{forw}
} 
Case (ii) is realized in the HERA detectors.
Here, the situation is much more involved.
One could make the assumption that nearly every one of the initial state
(LLA) photons is emitted collinearly,   
carries energy $(1-z)E_e$ and will hit the forward tagging device if its
geometrical size is big enough.
A simple estimate based on an analysis of the denominator of the
initial electron propagator after emission of the photon shows that
the tagging device must have a geometric size fulfilling the condition 
\ba
\Delta\theta_{\gamma} \gg 
\frac{m_e}{E_e} 
\ea
in order to collect most of these photons.
For an opening angle $\Delta\theta_{\gamma} = 0.5$ mrad and $E_e=25$
GeV this condition is fulfilled but not to such an extent that numerical
estimates of the LLA corrections would be sufficiently well.
In fact, the tagging device induces a cut on the photon angle and
leads to an additional logarithmic dependence of the cross-section on
it\footnote{ 
LLA corrections with photon tagging have been discussed in~\cite{jadach},
where a bremsstrahlung cross-section, reduced by tagged events, was estimated.
The aim was to reduce the radiative corrections from hard photon emission.   
Our modifications of the density function by photon cuts are quite similar to
the treatment in~\cite{jadach}.}.

We have to distinguish several contributions.
If both photons
hit the tagging device, i.e.:
\ba
0\leq \theta_{\gamma} \leq \Delta\theta_{\gamma},
\ea
the visible photon energy has to be defined as the sum
\ba
E^{\rm{vis,(iia)}}_{\gamma} =(1-z)E_e + E_{\gamma}.
\ea
Next, if the photon of the lowest-order scattering process does not
hit the photon tagger, i.e.:
\ba
\theta_{\gamma} > \Delta\theta_{\gamma},
\ea
only the collinear LLA photon contributes to
\ba
E^{\rm{vis,(iib)}}_{\gamma} =(1-z)E_e.
\ea
Finally, it may happen that the LLA photon does not hit the tagger, but the
lowest-order photon does:
\ba
0\leq \theta_{\gamma} \leq \Delta\theta_{\gamma},
\ea
and the visible photon energy is
\ba
E^{\rm{vis,(iic)}}_{\gamma} = E_{\gamma}.
\ea

\bigskip

All three cases contribute to the cross-section and have to be summed up
with the lowest-order cross-section:
\ba
 \frac{d^3\sigma_{(ii)}}
      {d\cos\theta^{'}_e dE^{'}_e d E^{\rm{vis}}_{\gamma}}
  &=&  
\frac{d^3\sigma_{\rm{brem}} }
      {d\cos\theta^{'}_e dE^{'}_e dE^{\rm{vis}}_{\gamma}}
+
 \frac{d^3\sigma_{(iia)}^{^{\rm{ini}}}} 
      {d\cos\theta^{'}_e dE^{'}_e dE^{\rm{vis}}_{\gamma}}
\nll \nll
&&+~
 \frac{d^3\sigma_{(iib)}^{^{\rm{ini}}}} 
      {d\cos\theta^{'}_e dE^{'}_e dE^{\rm{vis}}_{\gamma}}
+
 \frac{d^3\sigma_{(iic)}^{^{\rm{ini}}}} 
      {d\cos\theta^{'}_e dE^{'}_e dE^{\rm{vis}}_{\gamma}}.
\label{sigii}
\ea
We now have to extract $E^{\rm{vis}}_{\gamma}$ from~(\ref{generic}).
This will be done in the next three subsections 
with use of identity $\delta$-function integrations:
\ba
1 &\equiv& \int  d E^{\rm{vis}}_{\gamma}
\delta\left[ E^{\rm{vis}}_{\gamma} -(1-z)E_e - E_{\gamma}\right],
\label{delta2}
  \\
1 &\equiv& \int  d E^{\rm{vis}}_{\gamma}
\delta\left[ E^{\rm{vis}}_{\gamma} -(1-z)E_e \right],
\label{delta1}
  \\
1 &\equiv& \int  d E^{\rm{vis}}_{\gamma}
\delta\left[ E^{\rm{vis}}_{\gamma} - E_{\gamma} \right].
\label{delta3}
\ea
The three cases will be treated separately.

\subsubsection{Case (iia): Both photons hit the forward tagging device
\label{iia}
}
\begin{figure}[htbp]
\begin{center}
\mbox{
\epsfig{file=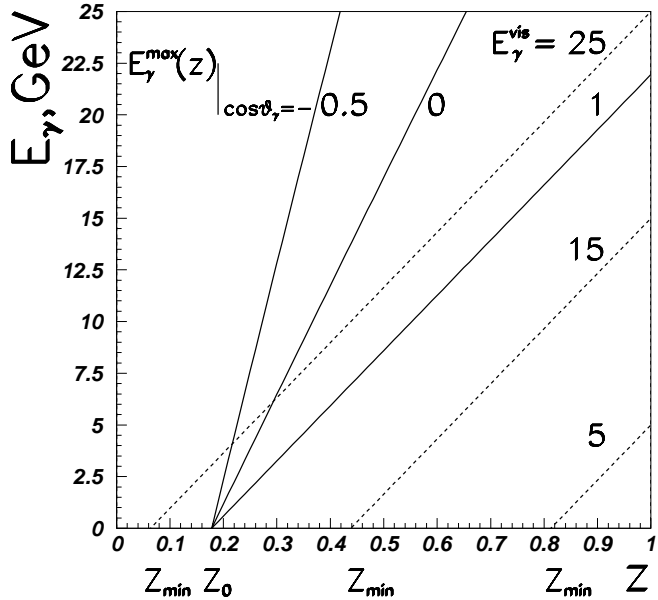,height=12cm,width=12cm,%
bbllx=0pt,bblly=0pt,bburx=730pt,bbury=730pt}}
\end{center}
\vspace{-3.0cm}
\caption{\it The physical region ($z,E_{\gamma}$) for Case (iia):
both photons hit the forward tagging device
\label{fiia}
}
\end{figure}

When two photons hit the forward tagging device, the cross-section gets:
\ba
 \frac{d^3\sigma_{(iia)}^{^{\rm{ini}}} }
      {d\cos\theta^{'}_e dE^{'}_e d E^{\rm{vis}}_{\gamma}}
&=&  
     \int_0^1 dz \, \rho_a(z) 
\nll
&& \times~ \Biggl\{\frac{d^2\sigma_{_{\rm{QED}}} }
      {d\cos\theta^{'}_e dE^{'}_e}
    \Biggl|_{E_e \rightarrow z E_e}   \Biggr.
    \Theta_z -
    \frac{d^2\sigma_{_{\rm{QED}}}}
      {d\cos\theta^{'}_e dE^{'}_e}    \Biggr\}
\delta\left[ E^{\rm{vis}}_{\gamma} -(1-z)E_e - E_{\gamma}\right] \nll
&=&  
     \int^{1}_{1-\frac{\Delta\theta^2_{\gamma}}{2}} d\cos\theta_{\gamma}
     \int^{2\pi}_{0}  d\varphi_{\gamma}
     \int_0^1 dz \rho_a(z) 
\nll
&&\times~ \Biggl\{\int^{{\hat E}^{\max}_{\gamma}}_0  d E_{\gamma}
\Biggl[\frac{d^5\sigma_{\rm{brem}}}
{d\cos\theta^{'}_e dE^{'}_e d\cos\theta_{\gamma} d\varphi_{\gamma}
  dE_{\gamma}} 
       \Theta(E_{\gamma} -{\bar E}_{\gamma})
\nll
&&  +~ 
   \frac{1}{4\pi}
      \frac{d^2\sigma_{_{\rm{VR}}}}{d\cos\theta^{'}_e dE^{'}_e}
\delta(E_{\gamma}) 
\Biggr] \Biggl|_{E_e \rightarrow z E_e} \Biggr. \Theta_z \nll
&&  
-~ 
\int^{E^{\max}_{\gamma}}_0  d E_{\gamma}
\Biggl[
      \frac{d^5\sigma_{\rm{brem}}}
{d\cos\theta^{'}_e dE^{'}_e d\cos\theta_{\gamma} d\varphi_{\gamma} dE_{\gamma}}
      \Theta(E_{\gamma} -{\bar E}_{\gamma})
\nll
&&  
+~ 
\frac{1}{4\pi}
      \frac{d^2\sigma_{_{\rm{VR}}}}{d\cos\theta^{'}_e dE^{'}_e}
\delta(E_{\gamma}) 
\Biggr] \Biggr\}
 \delta\left[ E^{\rm{vis}}_{\gamma} -(1-z)E_e - E_{\gamma}\right] .
\label{start1}
\ea
As mentioned, the density $\rho_a(z)$ will deviate from the inclusive
density~(\ref{lho}) due to the angular cut:
\ba 
\rho_a(z) = \frac{\alpha}{2\pi } \frac{1+z^2}{1-z} \left(
\ln \frac{m^2+E_e^2 \Delta\theta^2_{\gamma}
}{m^2} -1 \right) .
\label{lhoa}
\ea
After performing the integral over $z$ and some trivial re-orderings,
we arrive at
\ba
 \frac{d^3\sigma_{(iia)}^{^{\rm{ini}}} }
      {d\cos\theta^{'}_e dE^{'}_e d E^{\rm{vis}}_{\gamma}}
&=&
     \int^{1}_{1-\frac{\Delta\theta^2_{\gamma}}{2}} d\cos\theta_{\gamma}
     \int^{2\pi}_{0}  d\cos\varphi_{\gamma}
     \int^{{\min}\left\{E^{\rm{\rm{vis}}}_{\gamma}, E^{\max}_{\gamma}\right\}}
         _{{\max}\left\{{\bar E}_{\gamma},E^{\min}_{\gamma}\right\}}
 d E_{\gamma}
\nll
   &&\times~
\frac{\rho_a(z)}{E_e} 
\left|_{z=1-\frac{E^{\rm{vis}}_{\gamma}
-E_{\gamma}}{E_e}}\right. 
     \Biggl[
      \frac{d^5 \sigma_{\rm{brem}}}{d\COSTE d\EELP d\COSTG
d\varphi_{\gamma} d E_ {\gamma}}
    \Biggl|_{E_e \rightarrow z E_e}   \Biggr.
     \Theta_z
\nll
 &&  -~ \frac{d^5 \sigma_{\rm{brem}}}{d\COSTE d\EELP d\COSTG
d\varphi_{\gamma} d E_{\gamma}} 
     \Biggr]
\nll
&& + ~
\frac{\Delta \Omega_{\gamma}^{(iia)}}
{4\pi}
\frac{\rho_a(z)}{E_e} 
\left|_{{z_{\rm{min}}}=1-\frac{E^{\rm{vis}}_{\gamma}}{E_e}}\right. 
       \Biggl[
      \frac{d^2\sigma_{_{\rm{VR}}}}{ d\cos\theta^{'}_e dE^{'}_e }
    \Biggl|_{E_e \rightarrow {z_{\rm{min}}} E_e}   \Biggr.
       \Theta_{z_{\rm{min}}}
\nll
&&      -~
      \frac{d^2\sigma_{_{\rm{VR}}}}{ d\cos\theta^{'}_e dE^{'}_e  }
       \Biggr]\Theta(z_{\min} - z_0) .
\label{end1}
\ea
Everything is straightforward in the above derivation.
The only delicate point is the determination of the boundaries for the
integration over $E_{\gamma}$, and, related to this, the allowed range
for $E^{\rm{vis}}_{\gamma}$. 
This will be explained in some detail now with use of Figure~\ref{fiia}.
The integration over $E_{\gamma}$ proceeds along one of the dotted straight
lines with defining equation~(\ref{delta2}):  
\ba
E_{\gamma} = \left( E^{\rm{vis}}_{\gamma} - E_e \right) + z \, E_e .
\label{fiiaI}
\ea
The absolute maximum of $E^{\rm{vis}}_{\gamma}$ is reached for $z=1$,
and the absolute minimum for: 
\ba
z_{\min} = 1 - \frac{E^{\rm{vis}}_{\gamma}}{E_e},
\ea
where $E_{\gamma}$ vanishes.
For the hard photon integral in~(\ref{end1}) this minimum has
to be replaced by ${\bar E}_{\gamma}$ as discussed above.
The discussion of limits would be finished here if there wouldn't be
the scaled
kinematical upper limit for $E_{\gamma}$, ${\bar E}^{\max}_{\gamma}$,
which 
also has to be respected.
This limit depends on several variables: $z, E_e, E_p, E_e', \theta_{e}',
\varphi_{\gamma}, \theta_{\gamma}$.

\noindent
From~(\ref{emaxg}) one may see that the minimum value of
$E_{\gamma}^{\max}$, which is zero, is independent of the actual
values of the photonic angular variables. It depends, besides on $E_p$,
on $zE_e$ and external variables $E_e',\theta_e'$ and is reached
for fixed $E_e',\theta_e'$ at some value $z_0$ which 
may be easily derived from~(\ref{emaxg1}) (with neglected terms of
order ${\cal O}(m^2)$):
\ba
z_0 = z_0(E_e,E_p,E_e',\theta_e') &=& \frac{{\bar M}^2 -
  (p_1-k_2)^2}{2k_1(p_1-k_2)}. 
\label{zero}
\ea
As is visualized in Figure~\ref{fiia}, $z_0$ may be bigger or smaller
than $z_{\min}$ in dependence on $E^{\rm{vis}}_{\gamma}$.
If $z_0$ is smaller than $z_{\min}$, it is of no influence on the
integration region and the lower integration boundary is 
${\bar E}_{\gamma}$.
 In the other case, one has to study in detail
where the necessary condition $E^{\max}_{\gamma} >
E^{\rm{vis}}_{\gamma}$ is fulfilled. Some value $z_1$, corresponding
to some value $E_{\gamma}^{\min}$ may be found where this begins to be
fulfilled and the lower integration boundary becomes
$E_{\gamma}^{\min}$:
\ba
E_{\gamma}^{\min}
&=&
E_{\gamma}^{\max}(z_1E_e, E_p; E_e',\varphi_{\gamma},
\theta_{\gamma},\theta_e')
=E^{\rm{vis}}_{\gamma} - (1-z_1) E_e .
\ea
Therefore, $d^2\sigma_{_{\rm{VR}}}$ will not contribute to~(\ref{end1}).

We further have to study the fan of curves defining the
value of $E^{\max}_{\gamma}$, starting from
$z_0$ and depending on the parameters $\theta_{\gamma},
\varphi_{\gamma}$ as a function of $z$ at fixed $E_e',\theta_e'$.
Figure~\ref{fiia} shows that, for HERA kinematics, the curves are
nearly straight lines. 
If their curvature is strong enough, they may cross the straight
dotted integration path from above, in principle independent of the
relation between $z_0$ and  $z_{\min}$.
If this happens the upper integration limit in~(\ref{end1}) is not
$E^{\rm{vis}}_{\gamma}$ but ${\bar E}^{\max}_{\gamma}$.
The figure suggests that for the case of a forward tagging device
(with $\cos\theta_{\gamma} \approx 1$) the relevant solid curve is
nearly parallel to the dotted line. Then, for reasonable choices of
$E^{\rm{vis}}_{\gamma}$ no problems should occur from details of the
kinematics. 

To conclude the discussion, one has to integrate the scaled cross-section
over all values of
$E_{\gamma}$ on one of the dotted lines which are located below the
envelop of the fan of curves $E^{\max}_{\gamma}(z)$.
The unscaled cross-sections are integrated from ${\bar E}_{\gamma}$ to
$E^{\rm{vis}}_{\gamma}$.
In applications, we restrict ourselves to kinematics with $z_0 <
z_{\min}$ and $E^{\max}_{\gamma} > E^{\rm{vis}}_{\gamma}$.
The {\tt FORTRAN} program stops if $z_{\min} < z_0$.
\subsubsection{Case (iib): The LLA photon hits the forward tagging device
\label{iib}
}
\begin{figure}[bthp]
\begin{center}
\mbox{
\epsfig{file=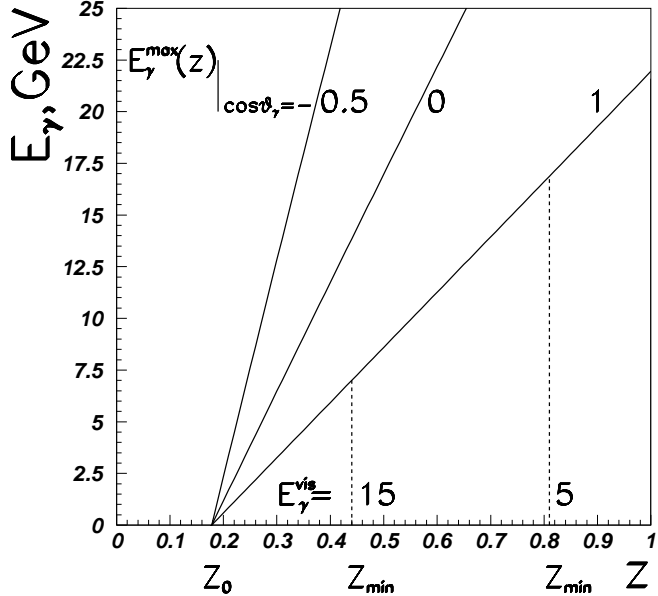,height=12cm,width=12cm,%
bbllx=0pt,bblly=0pt,bburx=730pt,bbury=730pt}}
\end{center}
\vspace{-3.0cm}
\caption{\it The physical region ($z,E_{\gamma}$) for Case (iib):
only the LLA photon hit the forward tagging device 
\label{fiib}
}
\end{figure}

If only the collinear initial-state LLA photon hits the forward
tagging device, the cross-section becomes:
\ba
 \frac{d^3\sigma_{(iib)}^{^{\rm{ini}}}}  
      {d\cos\theta^{'}_e dE^{'}_e d E^{\rm{vis}}_{\gamma}}
&=&  
     \int_0^1 dz \, \rho_b(z) 
\nll
&&\times~\Biggl\{ \frac{d^2\sigma_{_{\rm{QED}}} }
      {d\cos\theta^{'}_e dE^{'}_e}
    \Biggl|_{E_e \rightarrow z E_e}   \Biggr.
    \Theta_z -
    \frac{d^2\sigma_{_{\rm{QED}}}}
      {d\cos\theta^{'}_e dE^{'}_e}    \Biggr\}
\delta\left[ E^{\rm{vis}}_{\gamma} -(1-z)E_e\right] 
\nll
&=&  
     \int^{1-\frac{\Delta\theta^2_{\gamma}}{2}}_{-1} d\cos\theta_{\gamma}
     \int^{2\pi}_{0}  d\varphi_{\gamma}
     \int_0^1 dz \, \rho_b(z) 
\nll
&&\times~ 
\Biggl\{\int^{{\hat E}{\gamma}}_0  d E_{\gamma}
\Biggl[\frac{d^5\sigma_{\rm{brem}}}
{d\cos\theta^{'}_e dE^{'}_e d\cos\theta_{\gamma} d\varphi_{\gamma} dE_{\gamma}}
       \Theta(E_{\gamma} -{\bar E}_{\gamma})
\nll
&&  +    \frac{1}
{4\pi}
      \frac{d^2\sigma_{_{\rm{VR}}}}{d\cos\theta^{'}_e dE^{'}_e}
\delta(E_{\gamma}) 
\Biggr] \Biggl|_{E_e \rightarrow z E_e} \Biggr. \Theta_z 
\nll
&&  - \int^{E^{\max}_{\gamma}}_0  d E_{\gamma}
\Biggl[
      \frac{d^5\sigma_{\rm{brem}}}
{d\cos\theta^{'}_e dE^{'}_e d\cos\theta_{\gamma} d\varphi_{\gamma} dE_{\gamma}}
      \Theta(E_{\gamma} -{\bar E}_{\gamma})
\nll
&&  + \frac{1}{4\pi} 
      \frac{d^2\sigma_{_{\rm{VR}}}}{d\cos\theta^{'}_e dE^{'}_e}
\delta(E_{\gamma}) 
\Biggr] \Biggr\}
 \delta\left[ E^{\rm{vis}}_{\gamma} -(1-z)E_e\right]\;.
\label{start2}
\ea
The density function for the LLA correction is the same as that for
Case~(iia): 
\ba
\rho_b(z) = \rho_a(z).
\label{lhob}
\ea
The subsequent formal steps differ from Case (iia) since
$E_{\gamma}$
is not involved in the $\delta$-function. 
The limits of variation of
$E_{\gamma}$ are affected only by scaling:
\ba
 \frac{d^3\sigma_{(iib)}^{^{\rm{ini}}}}  
      {d\cos\theta^{'}_e dE^{'}_e d E^{\rm{vis}}_{\gamma}}
&=&  
\frac{\rho_a(z)}{E_e}  
\left|_{{z_{\rm{min}}}=1-\frac{E^{\rm{vis}}_{\gamma}}{E_e}}\right. 
   \Biggl\{
   \int^{1-\frac{\Delta\theta^2_{\gamma}}{2}}_{-1} d\cos\theta_{\gamma}
   \int^{2\pi}_{0}  d\cos\varphi_{\gamma}
\nll
   &&\times
   \int^{{\min}\left\{E^{\rm{res}}_{\gamma}, E^{\max}_{\gamma}\right\}}
       _{ {\bar E}_{\gamma} } d E_{\gamma}
   \Biggl[
   \frac{d^5 \sigma_{\rm{brem}}}{d\COSTE d\EELP d\COSTG
     d\varphi_{\gamma} d E_{\gamma}} 
   \Biggl|_{E_e \rightarrow {z_{\rm{min}}} E_e}   \Biggr.
   \Theta_{z_{\rm{min}}} 
\nll
  &&\hspace{3cm} -~
   \frac{d^5 \sigma_{\rm{brem}}}{d\COSTE d\EELP d\COSTG
     d\varphi_{\gamma} d E_{\gamma}} 
   \Biggr]
\nll
&& +~ \frac{\Delta \Omega_{\gamma}^{(iib)}}{4\pi} 
     \Biggl[
      \frac{d^2\sigma_{_{\rm{VR}}}}{ d\cos\theta^{'}_e dE^{'}_e }
     \Biggl|_{E_e \rightarrow {z_{\rm{min}}} E_e} \Biggr.
     \Theta_{z_{\rm{min}}} -
      \frac{d^2\sigma_{_{\rm{VR}}}}{ d\cos\theta^{'}_e dE^{'}_e  }
     \Biggr]
     \Biggr\} .
\label{end2}
\ea
The integration boundaries may be inspected in Figure~\ref{fiib}. 
One should notice that the upper integration limit is again
different  for the scaled and the unscaled cross-section integration. 
In~(\ref{end2}), the unscaled integration limits are 
used, however.
In the figure, they define the solid straight line ranging from the point
$(z,E_{\gamma}) = (1,{\bar E}_{\gamma})$  upwards until it meets
the actual value of $E_{\gamma}^{\max}$; the latter depends on the
kinematics.
The scaled limits are dependent on $z$ and define a dotted straight
line at fixed $z_{\min}$. 
These limits are automatically respected by means of
the factor $\Theta_{z_{\min}}$ in~(\ref{end2}).

Further, we introduced a new variable, $E^{\rm{res}}_{\gamma}$, in the
definition of the upper integration bound.
If somebody wants to impose a general cut on the maximum allowed
photon energy, it has to be done here and only here.

\subsubsection{Case (iic): The LLA photon does not hit the forward tagging
device 
\label{iic}
}
This case is very similar to Case~(i).
The correction may be read off from~(\ref{ini-i}) with a modified density
$\rho_c$ due to the cut on the angular integration for the LLA photon 
and with performing the angular integration for the Born photon as in
Case~(iia):  
\ba
 \frac{d^3\sigma_{(iic)}^{^{\rm{ini}}}} 
      {d\cos\theta^{'}_e dE^{'}_e d E^{\rm{vis}}_{\gamma}}
&=&
\int_{1-\frac{\Delta\theta^2_{\gamma}}{2}}^{1} 
d\cos\theta_{\gamma}
\int^{2\pi}_{0}  
d\varphi_{\gamma}
     \int_0^1 dz 
\rho_c(z)
\nll   
&& 
\Biggl\{
      \frac{d^5\sigma_{\rm{brem}} }
      {d\cos\theta^{'}_e dE^{'}_e d\cos\theta_{\gamma}
        d\varphi_{\gamma} 
dE_{\gamma}
}
    \Biggl|_{E_e \rightarrow z E_e}   \Biggr.
       \Theta_z
\nll   
&& 
-~   \frac{d^5\sigma_{\rm{brem}}}
{d\cos\theta^{'}_e dE^{'}_e d\cos\theta_{\gamma} d\varphi_{\gamma}
  dE_{\gamma}} 
\Biggr\} ,
\label{ini-c}
\ea
with
\ba 
\rho_c(z) = \frac{\alpha}{2\pi } \frac{1+z^2}{1-z}
\ln \frac{Q^2} {m^2+E_e^2 \Delta\theta^2_{\gamma}} .
\label{lhoc}
\ea
\section{Numerical Results and Discussion
\label{discussion}
}
\begin{figure}[htbp]
\begin{center}
\mbox{
\epsfig{file=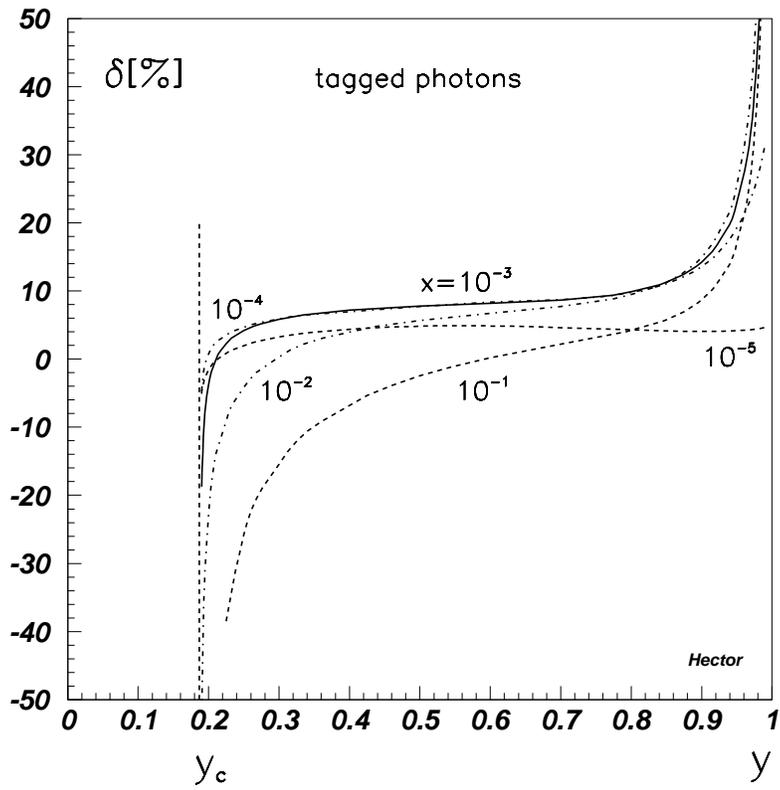,height=15.5cm,width=15.5cm,%
bbllx=0pt,bblly=0pt,bburx=730pt,bbury=730pt}}
\end{center}
\caption{\it 
Photonic correction 
$\delta = d^3\sigma^{\rm{ini}} / d^3\sigma_{\rm{brem}}
\times 100 \%$  for $E^{\rm{vis}}_{\gamma}=5$ GeV and
$\Delta\theta_{\gamma} = 0.5$ 
mrad for a forward tagging device
\label{fig7}}
\end{figure}

\begin{figure}[htbp]
\begin{center}
\mbox{
\epsfig{file=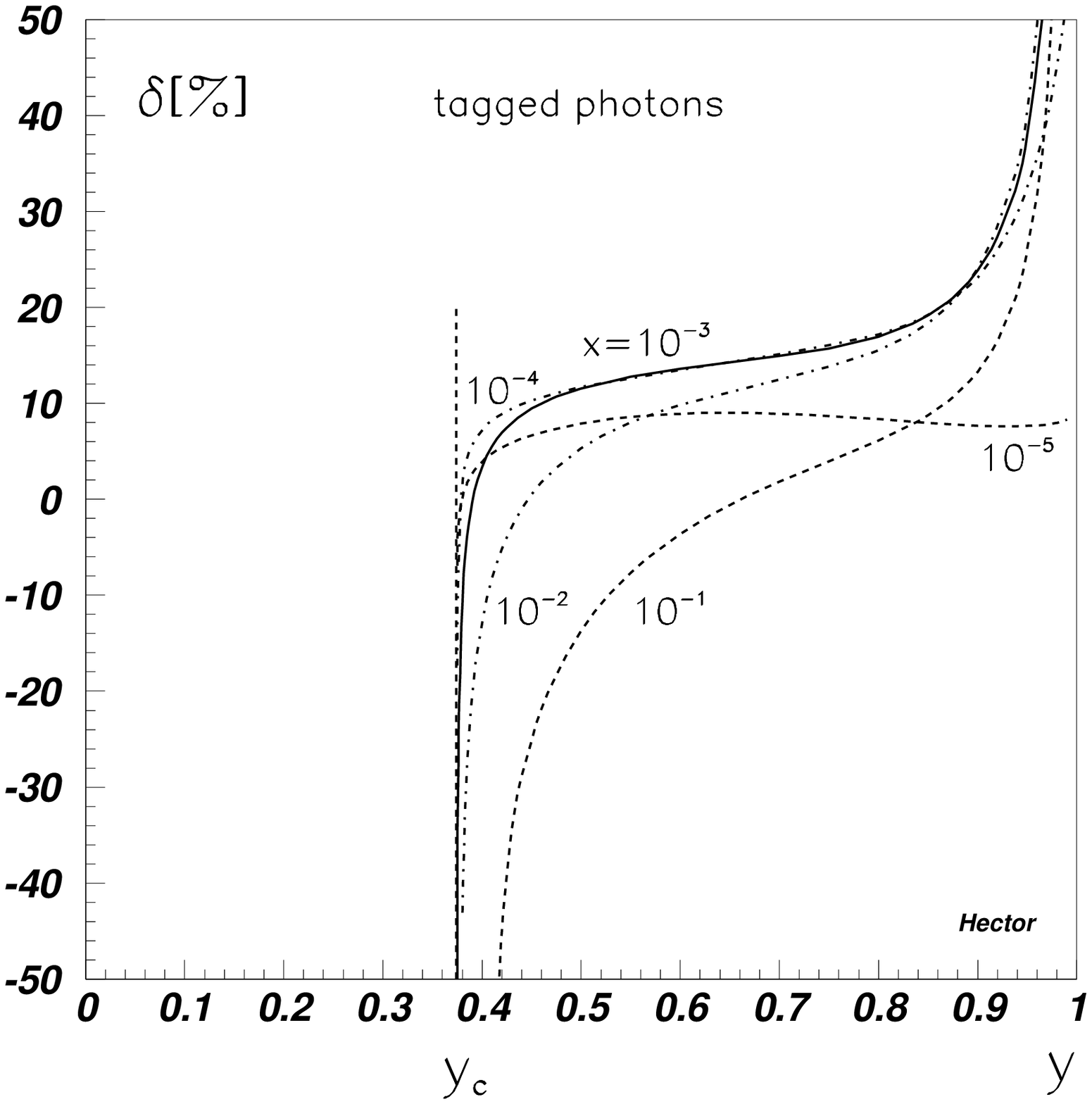,height=15.5cm,width=15.5cm,%
bbllx=0pt,bblly=0pt,bburx=730pt,bbury=730pt}}
\end{center}
\caption{\it 
The same as figure~6 but with $E^{\rm{vis}}_{\gamma}=10$ GeV
\label{fig7a}}
\end{figure}

Numerical results are shown for the H1 forward tagging device, which
is approximated here by 
a circle located symmetrically around the beam axis with
$\Delta\theta_{\gamma} = 0.5$ mrad.
The lowest-order cross-section has been shown in Section~\ref{low}.
The estimated radiative corrections $\delta$,
\ba
\delta = 
\frac {d^3\sigma^{\rm{ini}} / dx dy d E_{\gamma}^{\rm{vis}}}
{d^3\sigma_{\rm{brem}} / dx dy d E_{\gamma}^{\rm{vis}}} \times 100 \%,
\ea
are \hfill shown \hfill in \hfill Figures  \hfill \ref{fig7} and \ref{fig7a}. 
 \hfill
The \hfill $d^3\sigma^{\rm{brem}}$ \hfill is \hfill defined \hfill in
\hfill~(\ref{lowest}) \hfill and \hfill 
$d^3\sigma^{\rm{ini}}$ \hfill in \hfill (\ref{sigi}) \hfill or
\hfill (\ref{sigii}).

\noindent
For the figures, we consider Case~(ii) (forward tagging) and assume
$E^{\rm{vis}}_{\gamma}=5$ GeV and $E^{\rm{vis}}_{\gamma}=10$ GeV.  
The maximal allowed photon energy is assumed not to be restricted;
this corresponds to no photon observation outside the tagging device. 
As was mentioned already in the discussion of the Born cross-section,
there is a minimal value $y_c$ of $y$ which depends on the kinematics
($x, y, E_{\gamma}^{\rm{vis}}$).
Otherwise, the resulting corrections are similar to those known from
the inclusive measurements.
The Born cross-section dies out for $x > 10^{-4}$, thus limiting the
$Q^2$ values at HERA to be not bigger than several GeV$^2$.
For nearly all this kinematical range, the corrections stay well in
between about $\pm 20 \%$, mostly even within $0$ and $10 \%$.
Larger, negative QED corrections arise in the soft photon corner (at
small $y$) and large positive ones for hard radiation at high $y$.
The biggest, negative corrections are obtained at larger values of $x$
where the Born rates go down rapidly.
The soft photon corrections are weakened by higher-order contributions.
These may be taken into account by performing soft photon exponentiation. 
An approximation could be to do this for either the lowest-order
bremsstrahlung cross-section or the LLA density function separately.
An explicit higher-order calculation is needed for a sufficiently correct
answer. 
The numbers in the figures were produced without soft photon exponentiation.

\bigskip

In this article, we studied deep inelastic cross-sections at HERA energies with
photon tagging.
The lowest-order radiative process is calculated in ultra-relativistic
approximation with no additional restricting assumptions. 
The next-order corrections have been estimated with a
leading-logarithmic approximation neglecting the interferences between
diagrams with different topology concerning the two photons.
The resulting error is expected to be of next-to-leading order and could be
non-negligible for a forward tagging device.
Our calculations are performed strictly in order ${\cal O}(\alpha)$.
For soft photon emission, we know to overestimate the large, negative
corrections.
An exponentiation of soft photon corrections is recommended here if one wants
to improve the numerical results.
Again, problems due to interference effects remain unsolved so far.
For a forward tagger, the problem is not completely inclusive.
Logarithmic corrections due to the finite size of the tagger are
taken into account. 

Since the calculated corrections are not too big compared to the
typical size of photonic corrections in deep inelastic scattering
one may expect that they represent a quite good approximation of a
complete calculation of the order ${\cal O}(\alpha)$ photonic corrections.
    
\addcontentsline{toc}{section}{Acknowledgements}

\section*{Acknowledgements}
We are grateful to J.~Feltesse for drawing our interest to this problem.
We are indepted to A.~Arbuzov, A.~Glazov, and M.~Klein for fruitful
discussions, to 
S.~Jadach for directing our attention to reference~\cite{jadach}, 
and to H.~Anlauf, L. Favart, F.~Jegerlehner and M.~Klein  
for a careful reading of the manuscript and comments.

\clearpage
\appendix
\def\theequation{\Alph{section}.\arabic{equation}}
\section{Kinematics and Phase Space
\label{appA}
}
\ezero
The angular variables $\theta _e^{\prime}, \theta_{\gamma},
\varphi_{\gamma}$ are unrestricted.
The lower limits of $E_e^{\prime}$ and $E_{\gamma}$ are $m$ and zero,
respectively. 

The upper limits of the two energies depend on the
order of integration. 
We calculate the cross-sections in terms of the electron variables
$E_e^{\prime}, \theta 
_e^{\prime}$.
Therefore, we call them {\em external} variables and may derive the 
corresponding boundary from the condition
\ba
   W^2  &=& (k_1+p_1-k_2)^2 
\nll
 &=& (E_e +E_p -E^{'}_e)^2  - (p^{'}_e)^2
        -(p_e-p_p)^2 + 2 p^{'}_e (p_e-p_p) \COSTE
\nll
&\geq&  {\bar M}^2.
\label{w2}
\ea
Solving this inequality  for $E_e^{\prime}$ yields (\ref{starl}).

\begin{figure}[bthp]
\begin{center}
\mbox{
\epsfig{file=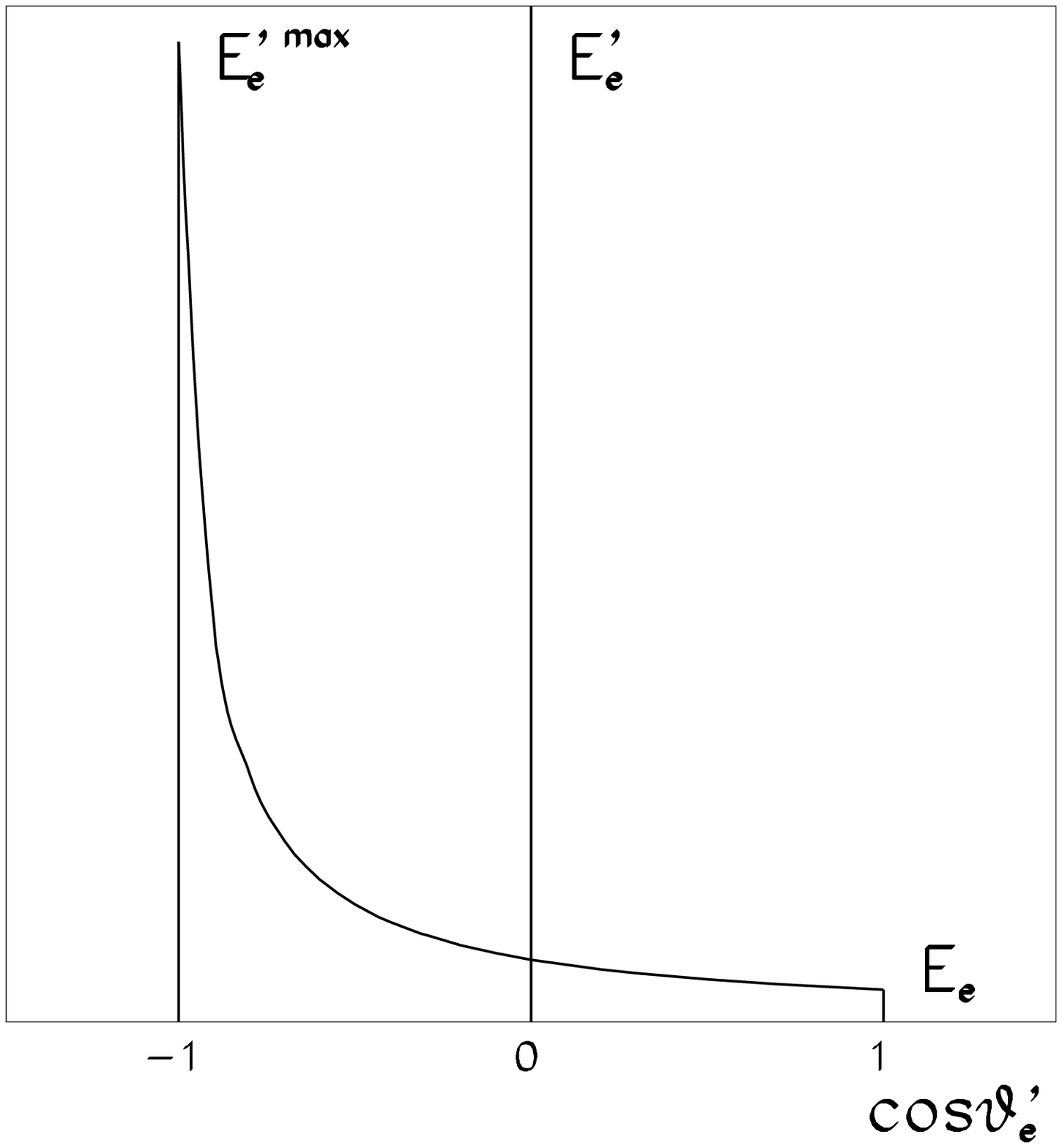,height=12cm,width=12cm,%
bbllx=0pt,bblly=0pt,bburx=730pt,bbury=730pt}}
\end{center}
\vspace{-3.0cm}
\caption{\it 
Kinematical boundaries for $\COSTE$ and $\EELP$
}
\label{mcs2}
\end{figure}

The two-dimensional region of variation of 
$\COSTE$ and $\EELP$ is shown in Figure~\ref{mcs2}.
Photonic variables are considered to be varying at fixed values of
electron variables and will be called {\em internal} variables.
The limit for $E_{\gamma}$  follows from 
a condition which relates $k_2$ and $k$ properly.
It is:
\ba
M_h^2  = (p_1+k_1-k_2-k)^2 \geq   {\bar M}^2.
\label{limit_eg}
\ea

In terms of natural variables:
\ba
M_h^2 &=& W^2   \\
 & & - 2  \EGAMMA \Biggl\{ \left[\PELP\COSTE-(\PEL-\PPR)\right]\COSTG
        +\PELP \SINTE \SINTG \COSPHI-\EELP+\EEL+\EPR  \Biggr\},
\nonumber
\label{mh2}
\ea

\noindent
{from} which we get
\ba
E^{\max}_{\gamma} = \frac{
W^2 - {\bar M}^2
}{
2\left[E_e +E_p 
                   -E^{'}_e+(p^{'}_e\cos\theta^{'}_e-[p_e-p_p])
\cos\theta_{\gamma}
                   +p^{'}_e\sin\theta^{'}_e
                           \sin\theta_{\gamma}
                           \cos\varphi_{\gamma}\right]
},
\label{emaxg1}
\ea
which in fact is~(\ref{emaxg}).

\begin{figure}[htbp]
\begin{center}
\mbox{
\epsfig{file=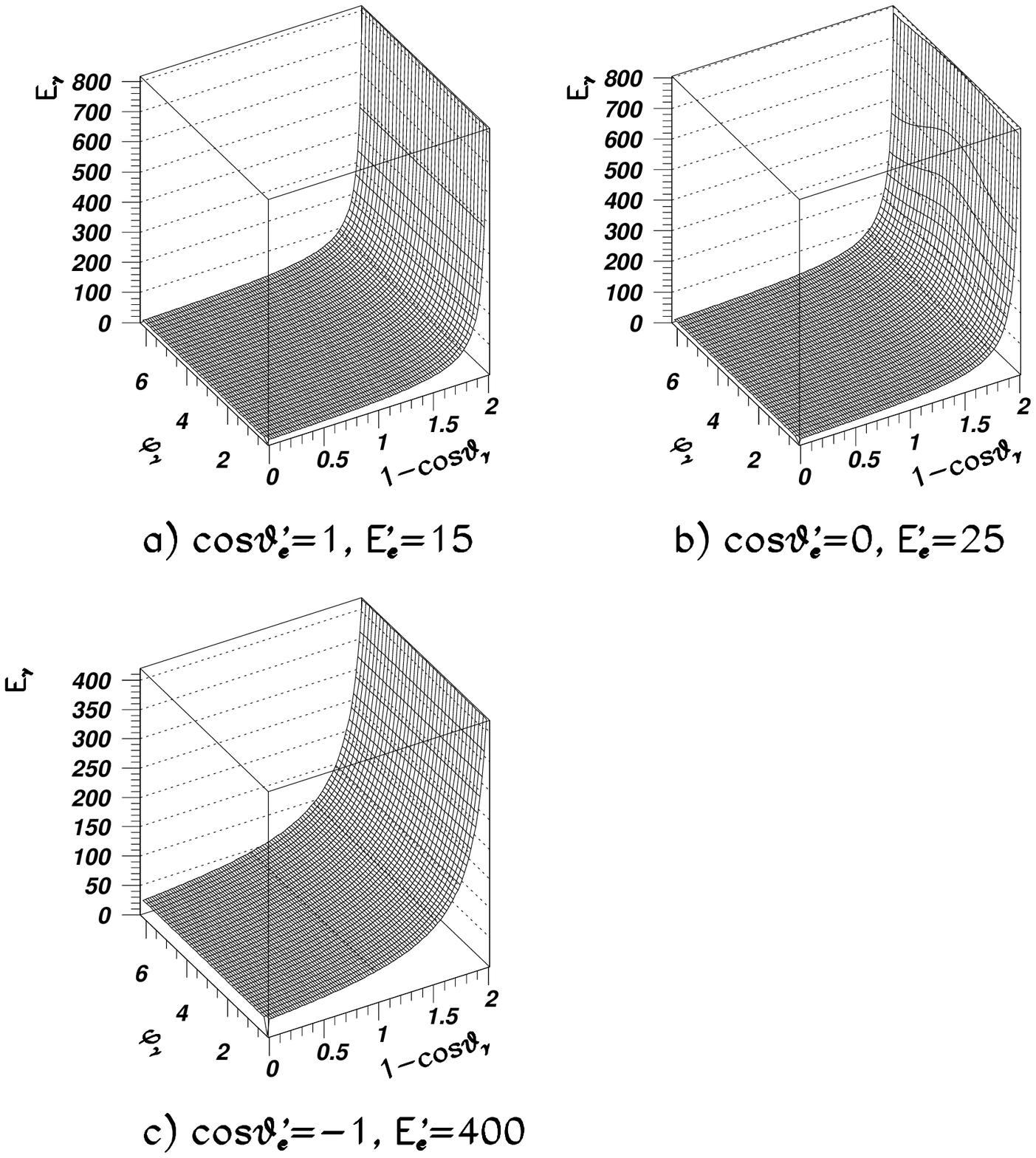,height=15.5cm,width=15.5cm,%
bbllx=0pt,bblly=0pt,bburx=580pt,bbury=580pt}}
\end{center}
\caption{\it 
Kinematical boundaries for $\cos\theta_{\gamma}, \varphi_{\gamma}, E_{\gamma}$
at given values of $E_e'$ and $\cos\theta_e'$
\label{3}
}
\end{figure}

In the ultra-relativistic approximation in $m$, (\ref{emaxg}) reduces 
for $ \theta_{\gamma} = 0 $
to 
\ba
E^{\max}_{\gamma}
               =  \frac
 {2 E_e E_p +(M^2-{\bar M}^2)/2
-  E_e E^{'}_e (1-\cos\theta^{'}_e)
-   E_p E^{'}_e (1+\cos\theta^{'}_e)  }
 {2 E_p - E^{'}_e(1-\cos\theta^{'}_e) }\;,
\ea
and for $ \theta_{\gamma} = \pi $ to
\ba
E^{\max}_{\gamma}
               =  \frac
 {2 E_e E_p +(M^2-{\bar M}^2)/2
-  E_e E^{'}_e (1-\cos\theta^{'}_e)
-   E_p E^{'}_e (1+\cos\theta^{'}_e)  }
 {2 E_e - E^{'}_e(1+\cos\theta^{'}_e) }\;,
\ea
These limiting cases are of use for an understanding of the  plots of
kinematical boundaries of variation 
of the {\it internal} variables
$\cos\theta_{\gamma},\varphi_{\gamma},E_{\gamma}$ which are shown in
Figure~\ref{3}. 

Next we have to discuss the phase space parameterization.
We aim at a check on the correctness of the kinematical boundaries
since the angular bounds have not been explicitly derived.
 
In Eq.~(I.3.9) we made use of a parameterization of the
phase space 
in terms of the invariants $ y_l, Q^2_l, y_h,$ $Q^2_h $, together with
one additional variable $z_2$ (or $z_1$):
\ba
 d\Gamma = \frac{\pi^2 S}{4 }
 d y_l \, d Q^2_l \, d y_h \, d Q^2_h \frac{dz_{1(2)}}{\pi \sqrt{R_z}}.
\label{gamie}
\ea
{From} explicit expressions for $y_l$ and $Q^2_l$ in terms of natural variables
it is easy to derive the Jacobean
\ba
  dy_l dQ^2_l = 2 p^{'}_e d\EELP d\COSTE .
\label{jacob}
\ea
{From}~({\ref{gamie}}) and~(\ref{jacob}) we get the doubly-differential
phase space in 
terms of $\COSTE,\EELP$ from a three-fold integration over invariants
with the aid of Eqs.~(I.D.2), (I.C.2), (I.C.3):
\ba
\frac{ d\Gamma                  }
     { p^{'}_e d\EELP d \COSTE  }
 = \frac{\pi^2 S}{2}
 \int_{{\bar M}^2}^{W^2} d M^2_h
 \int_{{Q^2_h}_{\min}}^{{Q^2_h}_{\max}}        d Q^2_h
 \int_{z_{\min}}^{z_{\max}} \frac{dz_{1(2)}}{\pi \sqrt{R_z}}
\label{dgammal}
=  \frac{\pi^2  (W^2 - M^2)^2 }{4 W^2},
\ea
where we also used~(I.3.7):
\ba
 d y_h \, d Q^2_h = \frac{1}{S}
  d M^2_h \, d Q^2_h. 
\ea
On the other hand, the phase space~(\ref{dgammal}) in terms of natural
variables was given in~(\ref{dgammall}):
\ba
\frac{ d\Gamma                  }
     { p^{'}_e d\EELP d \COSTE  }
 = \frac{\pi }{2}
 \int_0^{2\pi}d\PHI \int_{-1}^{+1} d\COSTG  \int_{0}^{E^{\max}_{\gamma}}
 E_{\gamma} dE_{\gamma}.
\label{dgammall1}
\ea
 We checked numerically that the phase space~(\ref{dgammal})
agrees within computer precision (14 digits) with
the new expression~(\ref{dgammall})
if one uses in~(\ref{dgammal}) the $W^2$ introduced in~(\ref{w2}).

Finally, we have to collect the integrals for the calculation of 
$\delta_{\rm{hard}}^{\rm{IR}}$:
 \ba
\left[ A \right] 
&=&
   \frac{1}{\pi}
\int_{\epsilon}^{\EGABAR} d E_{\gamma} E_{\gamma} \int_{-1}^{1} d \cos
\theta_{\gamma} \int_{0}^{2\pi} d \varphi_{\gamma} A.
 \ea
For sufficiently small $\EGABAR$ there is no dependence of the
angular integration limits on $E_{\gamma}$.
In this case, the integrations are straightforward and the results
are very simple:
\ba
\left[\frac{m^2}{z_1^2}\right] = \left[\frac{m^2}{z_2^2}\right] &=& \ln
\frac{\EGABAR}{\epsilon}, 
\\ \nll
\left[ \frac{1}{z_1z_2}\right]
&=& 
2  \lm  \ln \frac{\EGABAR}{\epsilon},
\ea
and $\lm$ is defined in~(\ref{lmlamm}).
The above two integrals are the equivalent of the tables of integrals
in appendix (I.D.3) for the present calculations.

\end{document}